\documentclass[preprint,showpacs,preprintnumbers,amsmath,amssymb]{revtex4}

\usepackage{graphicx}
\usepackage{dcolumn}
\usepackage{bm}
\usepackage{color}
\usepackage{booktabs}
\usepackage{multirow}

\usepackage{epstopdf}

\usepackage[version=4]{mhchem}

\AtBeginDocument{
\heavyrulewidth=.08em
\lightrulewidth=.05em
\cmidrulewidth=.03em
\belowrulesep=.65ex
\belowbottomsep=0pt
\aboverulesep=.4ex
\abovetopsep=0pt
\cmidrulesep=\doublerulesep
\cmidrulekern=.5em
\defaultaddspace=.5em
}

\usepackage{hyperref}

\begin{document}


\title{Production of rotationally cold methyl radicals in pulsed supersonic beams}

\author{Jonas Grzesiak}
\affiliation{Physikalisches Institut, Universit\"at Freiburg, 79104 Freiburg, Germany}
\author{Manish Vashishta}
\affiliation{Department of Chemistry, The University of British Columbia, Vancouver, British Columbia V6T 1Z1, Canada}
\author{Pavle Djuricanin}
\affiliation{Department of Chemistry, The University of British Columbia, Vancouver, British Columbia V6T 1Z1, Canada}
\author{Frank Stienkemeier}
\affiliation{Physikalisches Institut, Universit\"at Freiburg, 79104 Freiburg, Germany}
\author{Marcel Mudrich}
\affiliation{Department of Physics and Astronomy, Aarhus University, Ny Munkegade 120, 8000 Aarhus C, Denmark}
\author{Katrin Dulitz}
\affiliation{Physikalisches Institut, Universit\"at Freiburg, 79104 Freiburg, Germany}
\author{Takamasa Momose\footnote{Corresponding author: momose@chem.ubc.ca}}
\affiliation{Department of Chemistry, The University of British Columbia, Vancouver, British Columbia V6T 1Z1, Canada}

\date{\today}

\begin{abstract}
We present a comparison of two technically distinct methods for the generation of rotationally cold, pulsed supersonic beams of methyl radicals (\ce{CH3}): a plate discharge source operating in the glow regime, and a dielectric barrier discharge source (DBD). The results imply that the efficiency of both sources is comparable, and that molecular beams with similar translational and rotational temperatures are formed. Methane (\ce{CH4}) proved to be the most suitable radical precursor species. 

\end{abstract}

\pacs{Valid PACS appear here}
\maketitle

\section{\label{sec:Intro}Introduction}
The methyl radical, \ce{CH3}, is one of the smallest organic molecules and, owing to the presence of an unpaired electron, it is among the most important reaction intermediates in combustion and atmospheric chemistry \cite{Baulch:2005}. Therefore, detailed spectroscopic and reaction dynamics studies of this molecular radical have traditionally been of paramount interest to researchers \cite{Hirota1985, McFadden1972, Whitehead1996}. However, since radicals have to be produced \textit{in situ}, such studies are technically challenging owing to low signal intensities. Due to its planarity and the lack of a permanent electric dipole moment, the \ce{CH3} molecule does not show a pure rotational spectrum. In addition to that, there are no spectroscopic transitions in the visible range of the electromagnetic spectrum, and predissociation prevents the acquisition of well-resolved molecular spectra in the ultraviolet regime. Spectroscopic studies have thus been limited to rovibrational spectroscopy in the mid- and near-infrared region of the spectrum, as described in Refs. \cite{Hirota1985}, \cite{Davis:1997} and references therein. It is therefore desirable to produce cold and intense beams of \ce{CH3} in order to achieve good signal quality and long interrogation times with the sample. Using Zeeman deceleration, the production of translationally cold supersonic beams of \ce{CH3} has been demonstrated by some of us \cite{Momose2013}. Recently, we have also attained stationary samples of \ce{CH3} using magnetic trapping after Zeeman deceleration \cite{Momose:2017}.

A pulsed supersonic nozzle expansion, in which internal energy is converted into directed motion by the expansion of a highly pressurized gas into vacuum, is the most common approach to produce dense and internally cold molecular beams \cite{Miller1988, Morse1996}. Since radical production requires the dissociation of molecular bonds, external energy is applied during the expansion, which may lead to rotational and vibrational excitation of the molecules. A large number of techniques have been developed for the production of \ce{CH3} molecular beams, including pyrolysis \cite{Digiuseppe1982, Robinson1988, Whitehead1996}, photodissociation \cite{Zahedi1994, Holt1984} and discharge sources \cite{Ishiguro1996, Davis:1997}. However, only the pulsed discharge sources have been shown to yield cold molecular beams with rotational temperatures of around \mbox{25 K} \cite{Davis:1997, Ishiguro1996}. Ishiguro et al. used a supersonic jet expansion combined with a discharge modulation technique \cite{Ishiguro1996}, and Davis and co-workers have developed a shaped plate discharge source (slit discharge) to produce cold radical beams \cite{Anderson:1996, Davis:1997}.

In this article, we present the comparison of a plate discharge and a dielectric barrier discharge (DBD) source in combination with a home-built pulsed valve for the generation of rotationally cold \ce{CH3} radical beams. Both discharge techniques are widely used in industrial applications \cite{Kogelschatz:2003} and they have already been used for the production of molecular radicals and electronically excited atoms and molecules in supersonic jets \cite{Lewandowski:2004, Raunhardt:2008, Luria:2009, Ploenes:2016, vanBeek2001}. The two methods differ in terms of the underlying discharge mechanism and in terms of their technical complexity. The realization of a DBD is technically more demanding compared to the set-up of a plate discharge. A simple and robust two-plate electrode scheme can be used to ignite a DC discharge at relatively low voltages and at intermediate current strengths in the glow regime \cite{Roth:1995}. In this regime, the nozzle does not suffer from sputtering, and high electron excitation energies, which would lead to a heating of the gas pulse, are avoided. For the operation of a DBD, several AC high voltage pulses are applied to the electrodes, which are shielded from their surroundings using a dielectric material. This mode of operation ensures that the discharge current is kept at very low values \cite{Kogelschatz:2003, Luria:2009}. A DBD source has been shown to generate very cold and intense supersonic beams \cite{Luria:2009, Ploenes:2016}. Since the discharge is initiated in filaments, which are uniformly spread over a large surface, the formation of highly energetic species through arcing is prevented. It is therefore of particular interest to investigate whether a DBD can also be used as an efficient \ce{CH3} radical source.

In this paper, we give a detailed characterization and a direct comparison of both discharge sources in terms of \ce{CH3} radical intensities, beam velocity and rotational state distributions. We have also used two different precursor species, methane (\ce{CH4}) and di-tert-butyl peroxide (\ce{[(CH3)3CO]2}, DTBP), for radical generation. The relative efficiencies of both precursors in terms of \ce{CH3} radical production are discussed here as well. 

\section{\label{sec:Setup}Experimental setup}
\begin{figure}[ht!]
	\includegraphics[width=8.5cm]{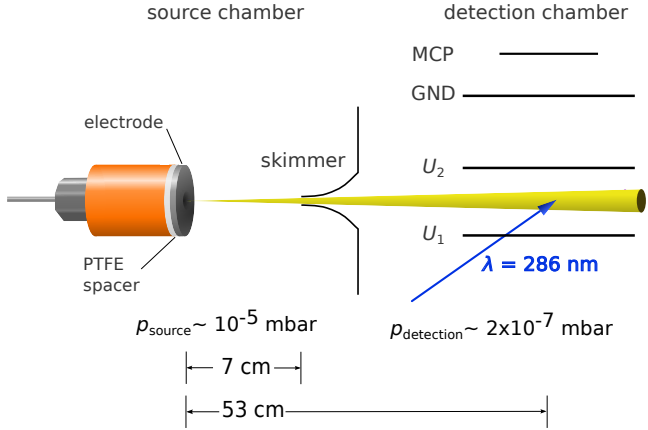}
	\caption{\label{fig:Setup} Schematic illustration of the experimental setup used for the characterization of the plate discharge source.}
\end{figure}

The experimental setup, which is schematically depicted in Fig.~\ref{fig:Setup} for the plate discharge source, consists of two differentially pumped vacuum chambers separated by a skimmer (2 mm diameter). The source chamber hosts a room-temperature, pulsed valve built at the Canadian Center for Research on Ultra-Cold Systems (hereafter referred to as the \textit{CRUCS valve}) which is optimized for the generation of short, variable pulse duration (25 - 100 $\mu$s) and intense gas pulses. The valve opening duration was set to 50 $\mu$s for all the experiments described in this paper, which generated gas pulses with a total duration of 90 - 100 $\mu$s at the discharge point. The valve design is conceptually similar to the commercially available Even-Lavie valve \cite{Even:2015}, i.e., a short current pulse generates a magnetic field inside a coil, which, in turn, lifts a magnetic plunger that admits gas from a high-pressure reservoir into the vacuum chamber. The nozzle is conically shaped at a 40$^{\circ}$ opening angle, and the valve orifice has a 250 $\mu$m diameter.

Methane (\ce{CH4}, Praxair, 99 \% purity, no further purification), or di-tert-butyl peroxide (\ce{[(CH3)3CO]2}, DTBP, Sigma Aldrich, 98 \% purity, no further purification) are used as radical precursor molecules, respectively. In the case of \ce{CH4}, the radical precursor species is supersonically expanded from a 30 \%  \ce{CH4}-\ce{Kr} premix in a gas cylinder at a 5 bar stagnation pressure. The precursor-noble-gas mixing ratio was initially optimized for the generation of methyl radical beams suitable for Zeeman deceleration \cite{Momose:2017} and is thus not further varied in the experiments described here. 
An $\approx0.6\%$ room-temperature gas mixture of DTBP in Kr carrier gas \cite{Indritz:1978} is obtained by filling a small amount of liquid DTBP into a lecture bottle and letting it equilibrate with Kr gas (5 bar overall pressure) overnight.

Molecular beams of \ce{CH3} radicals are formed by a supersonic expansion of the precursor gas mixture followed by bond cleavage initiated by one of the two discharge sources described below. Both of the sources are directly attached to the front plate of the valve (cf. Fig. ~\ref{fig:Setup}). The plate discharge source, which is schematically shown in Fig.~\ref{fig:Setup}, is based on the design described by Lewandowski et al. \cite{Lewandowski:2004}. It consists of one insulated, stainless steel electrode plate of the same outer diameter as the valve with a hole of 3 mm diameter and a thickness of 0.7 mm mounted to the valve head. The  electrode is set to high negative voltage, and the valve head serves as the ground electrode. The electrode is insulated against the valve using a polytetrafluoroethylene (PTFE) spacer (outer diameter as the valve, 7 mm hole diameter and 2.5 mm thickness). 

In our case, the supersonic expansion creates conditions of a steady charge flow that prevents the discharge from arcing \cite{Phelps:1960, Davis:1997}, and thus also avoids an unwanted heating of the supersonic beam. Such heating would be caused by the high current flow along very distinct and spatially small paths. We did not observe an increase of the beam temperature when the duration of the discharge pulse was increased. For this reason, we have applied DC voltages to the electrode during the experiments described here. Due to the voltage applied to the electrode, the direction of the gas flow is directed against the drag of the electrons. In contrast to schemes in which the electrode voltages are pulsed on and off for very short time periods \cite{Lewandowski:2004}, we found that the discharge process can be driven in the glow regime, i.e., without arcing, at relatively low DC voltages of around -1 kV. To ensure that the discharge is operated in the glow regime, we monitor the applied voltage using a voltage probe. In the case of arcing, a strong and rapid decrease of the voltage is observed. In contrast to that, a glow discharge is characterized by a much smaller and less abrupt voltage drop. In our setup, the use of an additional glow filament for the generation of seed electrons for the discharge, as reported in Refs. \cite{Lewandowski:2004} and \cite{Halfmann:2000}, was not required for the stable operation of the discharge unit. To optimize the plate discharge source, the amplitude of the electrode voltage was varied between -0.6 kV and -1.6 kV.

For comparison with the plate discharge source, we used a dielectric barrier discharge head (DBD) \cite{Even:2015} in combination with a CRUCS valve of the same dimensions as detailed for the plate discharge setup above. Apart from the pulsed valve assembly, the experimental setup is identical to the plate discharge setup. The alumina ($\mathrm{Al_2O_3}$) nozzle for the DBD  source (250 $\mu$m orifice diameter) has a 40$^{\circ}$ opening angle and a parabolic shape. The ring electrode inside the DBD head is driven by a ferrite-core step-up transformer built in-house which is optimized for the generation of peak-to-peak AC voltages up to 5 kV at a frequency of 1 MHz. The transformer is externally triggered by two externally programmed pulse trains from a commercial digital delay generator. In the experiments, a fixed number of 0.5 $\mu$s-long TTL pulses with a gap of \mbox{0.5 $\mu$s} is used which results in a time period of 1 $\mu$s for each channel. The TTL pulse trains of each channel are offset by \mbox{0.5 $\mu$s} with respect to each other. The external generation of the DBD pulse sequence allows for the optimization of the time delay between the AC voltage pulse for the DBD source and the pulsed valve, and it is used to adjust the duration of the AC voltage pulse. The operation of the discharge source is monitored by using a fast photodiode (Thorlabs, APD38-A) mounted to a window in the source chamber.
For the experimental characterization of the DBD source, both the peak-to-peak voltage, $U_{\mathrm{DBD}}$, and the number of AC voltage periods contained in a pulse train, $N_{\mathrm{P}}$, were adjusted. The discussion below is limited to $U_{\mathrm{DBD}}$ = 3.0 kV and 4.4 kV and to $N_{\mathrm{P}}=8$ and 120 for the following reasons: \ce{CH3} radical signal is only observed at $U_{\mathrm{DBD}} \geq$ 3.0 kV and $N_{\mathrm{P}} \geq$ 8, and arcing starts to occur at $U_{\mathrm{DBD}} >$ 4.4 kV which in turn leads to large fluctuations of the signal intensity. For $N_{\mathrm{P}} >$ 120, no further increase of the \ce{CH3} signal intensity was observed. In addition to that, for low $N_{\mathrm{P}}$, the delay of the pulse train with respect to the falling edge of the valve trigger was optimized to the maximum of the gas pulse. For $N_{\mathrm{P}} \geq$ 120, the DBD pulse train covered the full duration of the gas pulse. 

To detect \ce{CH3} radicals, [2+1]-resonance-enhanced multiphoton ionization (REMPI) of the molecule via the 4p Rydberg state ~\cite{Black:1988} is used in combination with mass-selected ion detection in a Wiley-McLaren-type ion-time-of-flight spectrometer. For this, laser light at 573 nm at a 10 Hz repetition rate is generated by a Nd:YAG-laser-pumped, pulsed dye laser (Sirah, PrecisionScan) and subsequently frequency-doubled in a BBO crystal (9 mJ pulse energy, 8 ns pulse duration) to yield the desired laser radiation at \mbox{286.3 nm}. The laser beam is focused into the interaction volume using a lens (350 mm focal length). The methyl ions are then accelerated towards an MCP detector perpendicular to the molecular beam axis by applying DC voltages of $U_1$ = 1200 V and $U_2$ = 600 V to the extraction plates, respectively. The use of DC extraction voltages prevents the detection of ions produced during the discharge process.

To characterize the discharge sources, rotationally resolved REMPI spectra and \ce{CH3} beam time-of-flight profiles (TOF) were measured (see below). Here, the time-of-flight is defined as the time between discharge excitation and laser ionization. The excitation time could be deduced from the sudden voltage decrease at the electrode (plate discharge source) and from the light pulses emitted by the discharge (DBD source), respectively.

To ensure the comparability of both discharge sources, the settings of both pulsed valves in the absence of a discharge are adjusted such that the beam temperatures and profiles of both valves are the same. This is done by monitoring the expansion of molecular oxygen (without a discharge) using a [2+1]-REMPI scheme from the X$^3\mathrm{\Sigma_g}$($v$=0) into the $\mathrm{\tilde{C}}^3\Pi_{\mathrm{g}}$($v$=2) state at 287.5 nm \cite{Russell:1987} (6 mJ pulse energy) prior to the characterization of each discharge source.

\section{Results and Discussion}
The experimental results for the two discharge sources presented in Sections \ref{sec:TOFs} and \ref{sec:ROT} below were obtained using \ce{CH4} as \ce{CH3} radical precursor. The efficiency of DTBP for \ce{CH3} production is discussed in Section \ref{sec:DTBP}.

\subsection{\label{sec:TOFs}Time-of-flight profiles}

\begin{figure}[ht!]
	\includegraphics[width=8.5cm]{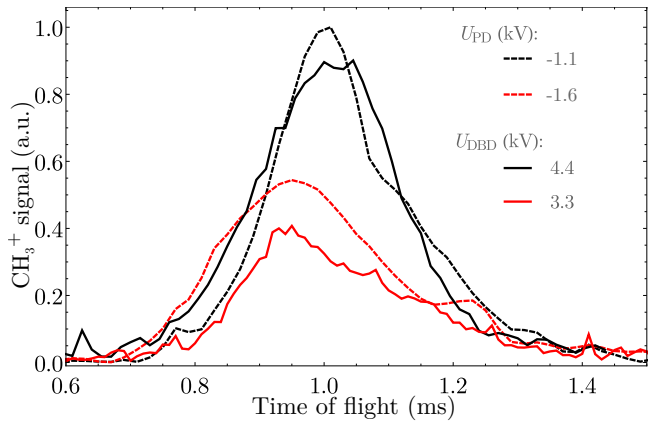}
	\caption{\label{fig:TOF} \ce{CH3} beam TOF traces at a two-photon wavenumber $\tilde{\nu}$ = \mbox{69852.8 cm$^{-1}$} for the plate discharge source (dashed lines) and the DBD source (solid lines) measured under different discharge conditions using \ce{CH4} as a radical precursor. The relative beam intensity is normalized to the maximum \ce{CH3+} ion yield obtained using the plate discharge source. A pulse train with $N_{\mathrm{P}}=120$ is used for the DBD source. For discharges using \ce{CH4} as a radical precursor, \ce{CH3+} ion signal is not observed when the discharge voltage is set to zero.} 
\end{figure}

\ce{CH3} beam TOF traces for both discharge methods are shown in Fig. \ref{fig:TOF}. For clarity, only a selection of traces is given. As can be seen from Fig. \ref{fig:TOF}, similar \ce{CH3} radical intensities can be observed for both discharge sources under certain experimental settings. For the plate discharge, we have found that radical production sets in at $U_{\mathrm{PD}} \approx$ -0.6 kV, and a maximum in signal intensity is measured at around -1.1 kV. Above this voltage, a decrease in signal intensity and a significant broadening of the TOF spectrum -- which is indicative of a significant heating of the beam by the discharge -- is observed. As the voltage is increased, the maximum of the TOF profile is shifted to shorter times, i.e. the beam velocity is increased.

\begin{figure}[ht!]
	\includegraphics[width=8.5cm]{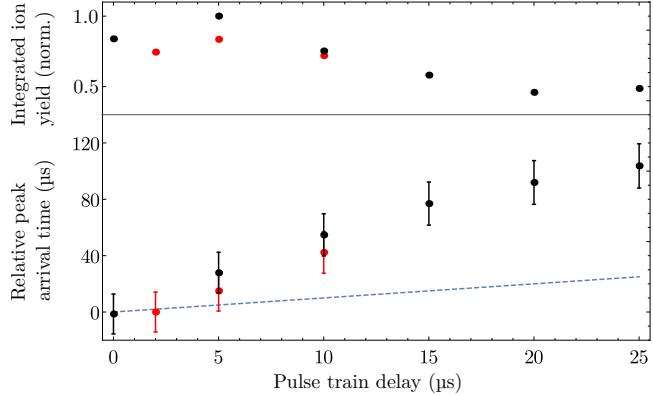}
	\caption{\label{fig:Delay} Integrated \ce{CH3+} ion yield (upper panel) and time shift of the TOF maximum (lower panel) as a function of the DBD pulse train delay for $N_{\mathrm{P}}=8$ (red points) and $N_{\mathrm{P}}=15$ (black points). Here, the time delay is defined as the time difference between the falling edge of the valve trigger pulse and the starting time of the DBD pulse sequence. The dashed line is a guide to the eye (see main text). The uncertainty is given as the error in the TOF arrival time only.}
\end{figure}

For the DBD source, the $\mathrm{CH_3}$ signal intensity is increased as the applied voltage $U_{\mathrm{DBD}}$ is raised. The highest $\mathrm{CH_3}$ radical yield is thus observed at the highest voltage $U_{\mathrm{DBD}}$ that can be applied without causing arcing, and for long pulse trains. At lower values of $U_{\mathrm{DBD}}$, a shift of the mean velocity to slightly higher values is observed. This can be due to an asymmetric density profile of the gas pulse, i.e. due to a higher intensity of the molecular beam at earlier times. Using very short pulses with $N_{\mathrm{P}}=8$ or $N_{\mathrm{P}}=15$, we have also observed a dependency of the peak arrival times upon the time delay for the DBD pulse train with respect to the valve trigger timing (cf. Fig. ~\ref{fig:Delay}). As can be seen from the lower panel in Fig. ~\ref{fig:Delay}, the mean arrival time of the molecules in the detection volume is not equal to the change in pulse train delay (dashed line). Instead, the peak arrival time is delayed much further than what would be expected from the set pulse train delay, i.e. a faster (slower) beam is produced at early (late) pulse train delays. The early fraction of the gas pulse is probably faster than the later fraction, because it is less efficiently cooled by collisions during the expansion process. The \ce{CH3+} ion yield obtained at each DBD pulse train delay (upper panel in Fig. ~\ref{fig:Delay}) reflects the intensity profile of the emitted gas pulse. Our findings also show that \ce{CH3} radical production is induced during the expansion process, where several velocity classes can be addressed during a very short time interval. Hence, to obtain a relatively slow supersonic beam, which is, for example, preferably for supersonic beam deceleration techniques, DBD excitation at late pulse train delays is preferable. We have also observed a similar, trigger-delay-dependent time shift of the TOF profiles for a pulsed plate discharge source (not shown).
To determine the beam velocity, each TOF profile was fitted to a shifted Boltzmann velocity distribution of a supersonic expansion which was convoluted with a rectangular function. For simplicity, we assume that the radicals are produced within a rectangular pulse shape of width 50 $\mu$s which is equal to the FWHM of the observed voltage drop during the discharge process. Since the discharge duration is small compared to the flight time of the beam through the apparatus, the uncertainty related to this approximated discharge profile is small but not negligible.
The molecular density in the expanding beam is only sufficient to induce a discharge very close to the nozzle orifice. Since this distance is small compared to the overall distance between nozzle and detector, the uncertainty related to the flight distance is also small. We estimate that the overall uncertainty of the velocity determination is within 3$\%$ for the mean velocity of the beam. A relative comparison of the velocities under different excitation conditions is thus possible. However, absolute values of the beam velocities cannot be accurately given due to the simple rectangular pulse shape assumed for the excitation process. The uncertainty for the full width at half maximum (FWHM) is about 10$\%$, which is due to the uncertainties in the fit procedure, i.e. a higher (lower) FWHM can be balanced by a higher (lower) amplitude and then still yield a reasonable fit to the data. 

The measured values of the mean \ce{CH3} beam velocity, the FWHM of the velocity distributions and the relative signal intensities obtained from the TOF profiles in \mbox{Fig. \ref{fig:TOF}} are summarized in Table \ref{tav:FWHM_PD} for the plate discharge source and in Table \ref{tav:FWHM_DBD} for the DBD source. As can be seen from \mbox{Table \ref{tav:FWHM_PD}}, both the beam velocity and the FWHM increase as the applied voltage is raised. In terms of the longitudinal beam temperature, the broadening of the TOF profile between -0.65 kV $\leq U_{\mathrm{PD}} \leq$ -1.6 kV corresponds to a temperature increase from 2 K to 12 K, i.e., by a factor of six. However, the maximum signal intensity is observed at $U_{\mathrm{PD}}$ = -1.1 kV. Hence, it has to be decided whether a beam with a high signal intensity and a larger velocity spread or a beam with a lower signal intensity and a smaller velocity spread is desired. In our Zeeman deceleration measurements, a voltage of $U_{\mathrm{PD}}$ = -1.1 kV is chosen for the electrode to maximize the radical density; the small increase in beam velocity can be compensated for by using a higher phase angle for deceleration \cite{Momose:2017}.
\begin{table}[ht!]
	\caption{\label{tav:FWHM_PD}Summary of \ce{CH3} beam characteristics obtained using the plate discharge source.}
	\begin{tabular}{rccc}
		\toprule
		& \multicolumn{3}{c}{$U_{\mathrm{PD}}$ (kV)}\\
		& -0.65 & -1.1 & -1.6 \\
		\midrule
		Mean beam velocity (m/s) & 510 & 540 & 560\\
		FWHM (\%) & 17 & 24 & 35\\         
		Relative beam intensity (a.u.) & 0.60 & 1 & 0.55\\
		\bottomrule
	\end{tabular}
\end{table}
\begin{table}[ht!]
\caption{\label{tav:FWHM_DBD}  Summary of \ce{CH3} beam characteristics for the DBD source. The beam intensity is normalized to the maximum \ce{CH3+} ion yield obtained using the plate discharge source (cf. Table \ref{tav:FWHM_PD}).}
\begin{tabular}{rcc}
\toprule
& $U_{\mathrm{DBD}}$ (kV)  &  $N_{\mathrm{P}}=120$ \\
\midrule
Mean beam velocity (m/s) & 3.3    & 555     \\
                         & 4.4    & 550     \\
FWHM (\%)                & 3.3    & 30     \\
                         & 4.4    & 28      \\
Relative beam intensity (a.u.)        & 3.3   & 0.33    \\
                                      & 4.4   & 0.87    \\
\bottomrule
\end{tabular}
\end{table}

\subsection{\label{sec:ROT}Rotationally resolved REMPI spectra}
\begin{figure}[ht!]
	\includegraphics[width=8.5cm]{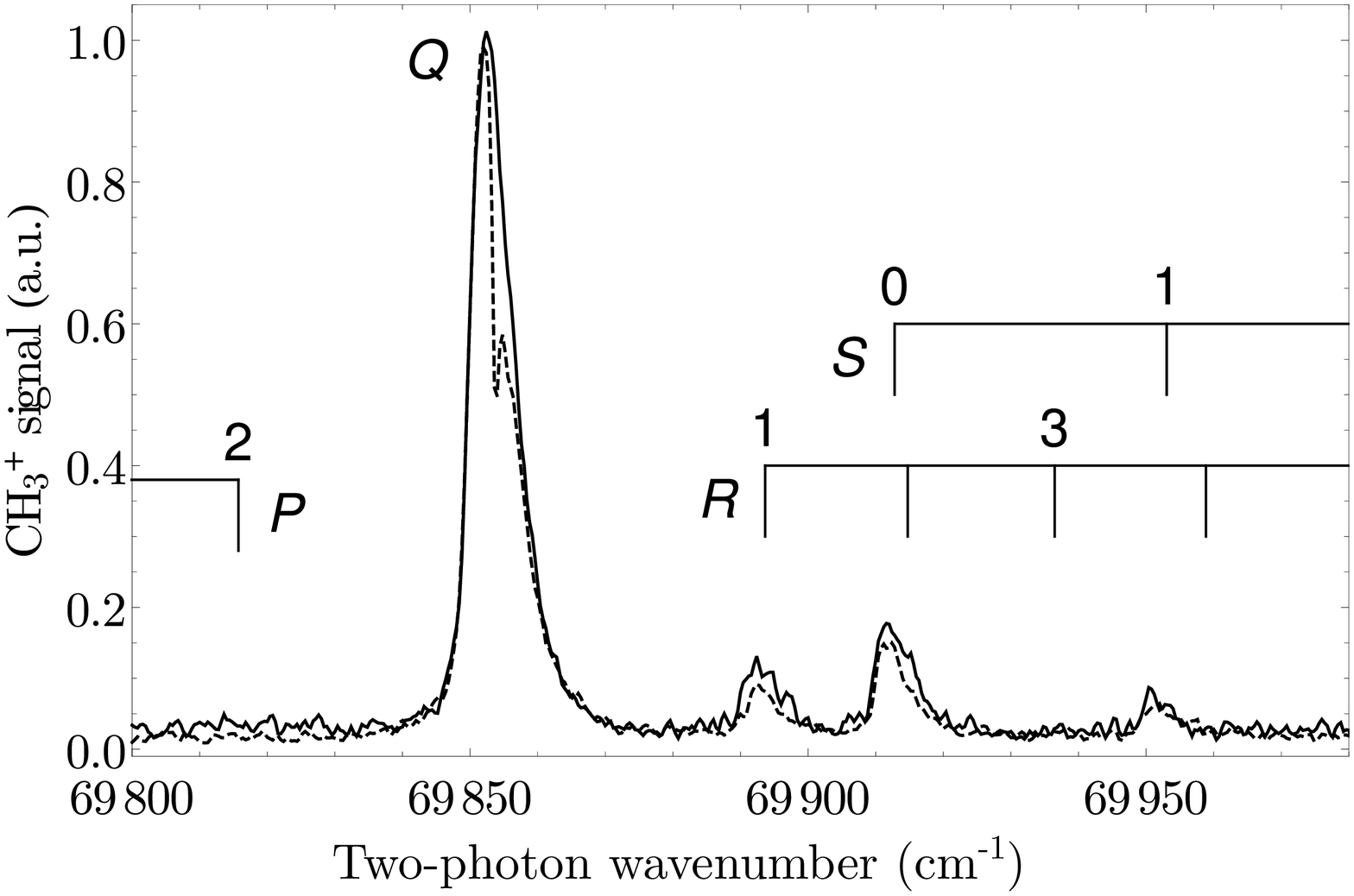}
	\caption{\label{fig:Rot_DBD} Rotationally resolved REMPI spectra of \ce{CH3} obtained using the plate discharge source (dashed line) and the DBD source (solid line). For the plate discharge source, $U_{\mathrm{PD}} = -1.1$ kV was used. For the DBD source, a voltage of $U_{\mathrm{DBD}}= 4.4$ kV was applied and a pulse train with $N_{\mathrm{P}}= 120$ was used. All traces are normalized to the maximum signal intensity at $\tilde{\nu}$ = 69852.8 cm$^{-1}$ (Q branch). Spectroscopic assignments are given on top of the spectra, where the labels P, Q, R and S correspond to $\Delta N = N^{\prime}-N^{\prime\prime}$ = -1, 0, 1, and 2, respectively. The dip in the plate discharge spectrum at $\tilde{\nu} \approx$ 69855 cm$^{-1}$ is an experimental artifact caused by instabilities in the laser system. The linewidth of the REMPI signal is limited by the lifetime of the intermediate 4p Rydberg state of \ce{CH3} \cite{Black:1988}.}
\end{figure}

Fig. \ref{fig:Rot_DBD} shows rotationally resolved REMPI spectra of \ce{CH3} obtained using the plate discharge and the DBD source, respectively. Similar spectra were obtained under all experimental conditions, but for reasons of clarity, only two exemplary spectra are depicted. The experimental conditions were the same as for the measurement of TOF spectra (Section \ref{sec:TOFs}). Here, $N^{\prime\prime}$ and $N^{\prime}$ denote the rotational angular momenta of the electronic ground state and of the electronically excited state, respectively, and $K^{\prime\prime}$ and $K^{\prime}$ are the corresponding projections of $N^{\prime\prime}$ and $N^{\prime}$ onto the principal axis. The spectroscopic assignments are labeled as P, Q, R and S and correspond to transitions with $\Delta N = N^{\prime}-N^{\prime\prime}$ = -1, 0, 1, and 2, respectively. As can be seen from the spectra in Fig. \ref{fig:Rot_DBD}, only transitions arising from $N^{\prime\prime}=0$ (S(0)) and $N^{\prime\prime}=1$ (R(1) and S(1)) have non-zero spectral intensity regardless of the discharge source\footnote{Since P(2) has zero spectral intensity, R(2) must also be zero, so that the transition at $\tilde{\nu}$ = 69912 cm$^{-1}$ can be unambiguously assigned to S(0).}. We can thus conclude that the discharge process does not affect the rotational cooling of \ce{CH3} into the lowest-lying rotational state of each nuclear-spin isomer, ortho-\ce{CH3} ($N^{\prime\prime}=0, K^{\prime\prime}=0$) and para-\ce{CH3} ($N^{\prime\prime}=1, |K^{\prime\prime}|=1$), by the supersonic expansion. Judging from the energy-level structure of the molecule, we can deduce a rotational temperature of $\leq$ 15 K, which is colder than the results obtained in Refs. \cite{Davis:1997} and \cite{Ishiguro1996}. This means that the heating of the supersonic beam at high discharge voltages only affects the translational motion and thus the longitudinal beam temperature of the molecules in the supersonic beam. This finding is of particular interest for applications in spectroscopy and supersonic beam deceleration experiments, where not only the velocity distribution of the beam is important but also the population of internal states plays a crucial role.



\subsection{\label{sec:DTBP}DTBP as a radical precursor}
It is known that, upon heating, di-tert-butyl peroxide (\ce{[(CH3)3CO]2}, DTBP) decomposes into two acetone molecules and two methyl radicals \cite{Raley1948}. The activation energy for unimolecular decomposition is very low (1.64 eV \cite{Dickey1949}), so that even a mild discharge should be sufficient to produce a considerable amount of \ce{CH3} radicals. In contrast to other peroxides, DTBP does not react with metal surfaces \cite{Dickey1949} and it is thus suitable for use with pulsed valves. DTBP has, thus far, been mainly used for \ce{CH3} radical generation in flash pyrolysis sources \cite{Yamada1981, Digiuseppe1981, Digiuseppe1982, Hudgens1983, Hoffmann1985, Robinson1988, Balucani2011}. A 60 Hz AC discharge of DTBP has also been tested for \ce{CH3} production in a gas cell \cite{Yamada1981, Amano1982} but not in a supersonic jet. Here, we explore the possibility of using DTBP in combination with a discharge source attached to a pulsed valve.

\begin{figure*}[htb!]
	\includegraphics[width=0.97\textwidth]{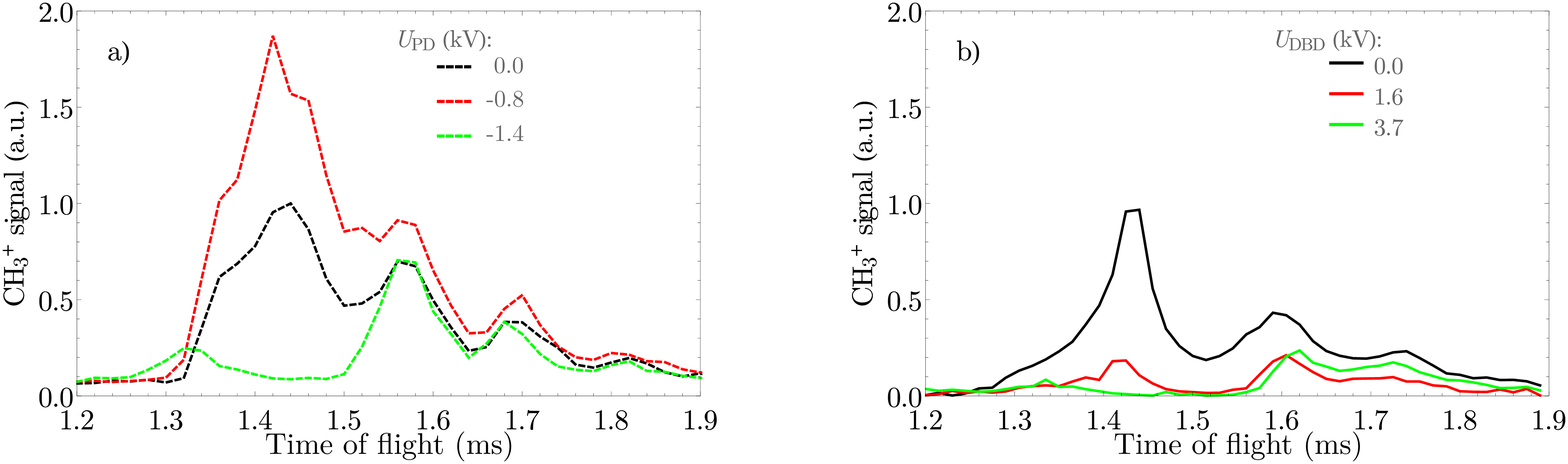}
	\caption{\label{fig:tBu_DBD} \ce{CH3} beam TOF traces at $\tilde{\nu}$ = 69852.8 cm$^{-1}$ for a) the plate discharge source and b) the DBD source using DTBP as a radical precursor. The nozzle conditions were optimized independently for both the plate discharge and the DBD source in order to obtain the maximum signal difference between discharge on and off. The relative beam intensity in both figures is normalized to the maximum \ce{CH3+} ion yield measured when the discharge is turned off (black curves). In a), traces at different electrode voltages are shown (green and red curves). In b), plots at distinct peak-to-peak voltages are displayed (green and red curves, at $N_{\mathrm{P}}$ = 120). The dissociative ionization of DTBP by the detection laser pulse also leads to the formation of \ce{CH3+} even when the discharge is turned off.}
\end{figure*}
\begin{figure*}[htb!]
	\includegraphics[width=8.5cm]{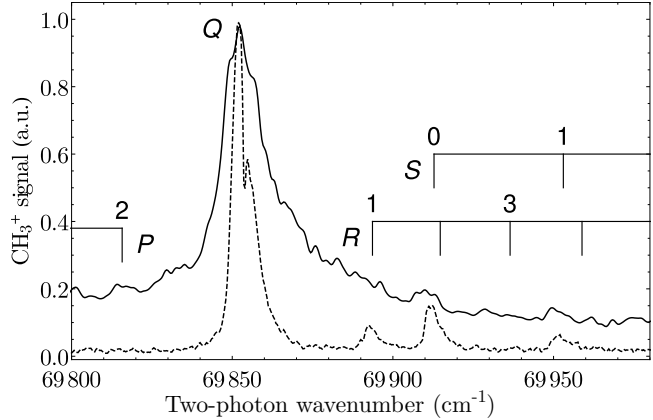}
	\caption{\label{fig:REMPI_DTBP} Rotationally resolved REMPI spectra of \ce{CH3} radicals obtained using DTBP (solid line) and \ce{CH4} (dashed line) as precursors. A plate discharge source was used which was operated at a voltage of $U_{\mathrm{PD}}$ = -0.7 kV for DTBP and at $U_{\mathrm{PD}}$ = -1.1 kV for \ce{CH4}, respectively. All traces are normalized to the maximum signal intensity at $\tilde{\nu}$ = 69852.8 cm$^{-1}$ (Q branch). Spectroscopic assignments are the same as in Fig. \ref{fig:Rot_DBD}.}
\end{figure*}

\ce{CH3} radical TOF traces obtained from DTBP discharges are shown in Fig. \ref{fig:tBu_DBD}. Even in the absence of a discharge, \ce{CH3+} ion signal is observed (black curves) which is probably due to the photodissociation of DTBP and subsequent ionization of thus produced \ce{CH3} radicals by the detection laser. The longer flight times through the apparatus compared to discharges in \ce{CH4} (cf. \mbox{Fig. \ref{fig:TOF}}) are due to the larger amount of Kr carrier gas as well as the higher mass of the DTBP precursor species. The observation of several TOF peaks is caused by a rebouncing of the valve plunger at the long valve opening times, which were used to obtain sufficient methyl radical signal intensity. For the plate discharge source, the \ce{CH3+} signal intensity at a flight time of around 1.4 ms is increased when the discharge is operated at $U_{\mathrm{PD}}$ = -0.8 kV (red curve in Fig. \ref{fig:tBu_DBD} a)) which indicates that \ce{CH3} radicals are indeed produced during the discharge process. However, the \ce{CH3+} signal intensity originating from the discharge is only 30 \% of the intensity obtained using the plate discharge source. At the trailing edge of the gas pulse (flight times $\geq$ 1.5 ms), the gas density was probably not high enough to maintain the discharge to produce a large number of \ce{CH3} radicals to detect. At high DC voltages ($U_{\mathrm{PD}}$ = -1.4 kV, green curve in Fig. \ref{fig:tBu_DBD} a)), the \ce{CH3+} signal at flight times around 1.4 ms is nearly depleted. This suggests that, at high DC voltages, DTBP rapidly decomposes into atomic and molecular fragments other than \ce{CH3}. In addition to that, the measured rotationally resolved REMPI spectrum (Fig. \ref{fig:REMPI_DTBP}) displays a much broader Q branch (FWHM of 20 cm$^{-1}$) compared to the spectrum obtained using \ce{CH4} as radical precursor (FWHM of 6 cm$^{-1}$). Furthermore, the P(2) transition can also be observed for a DTBP discharge. These observations indicate a higher rotational temperature for \ce{CH3} radicals obtained from DTBP in comparison to using \ce{CH4} as a precursor in the plate discharge source.
Using the DBD source, we were unable to see an increased \ce{CH3+} signal intensity upon discharge operation. Again, we attribute this to radical decomposition, since the lowest value of $U_{\mathrm{DBD}}$ is already higher than the maximum voltage for $U_{\mathrm{PD}}$ used for the plate discharge source (cf. Fig. \ref{fig:tBu_DBD} a) and b)). Methyl radicals may still be produced at very low AC discharge voltages which would -- in our case -- require a redesign of the step-up transformer core and could thus not be studied here.

\subsection{Conclusion}
We have found that both a plate discharge and a DBD source can yield cold supersonic beams of methyl radicals of similar intensity. Even though the absolute values of the mean velocities cannot be determined with high accuracy, the relative velocity obtained from the TOF profiles indicates that both discharge sources lead to an increase of the longitudinal beam temperature, and this effect is highest at high electrode voltages. In contrast to that, the rotational cooling of the beam into the lowest-lying rotational state of each nuclear-spin isomer, induced by inelastic collisions during the supersonic expansion, is not affected by the discharge process. However, in terms of technical complexity, a plate discharge source is much easier to set-up compared to a DBD source. In particular, a plate discharge source does not require special ceramic and magnetic discs, and additional electronics equipment is not necessary for the generation of a pulse train. The use of \ce{CH4} as a radical precursor is preferable to DTBP, since it provides a more efficient, rotationally cold and background-free source of \ce{CH3} for both a plate discharge and a DBD setup at various different experimental settings. In our laboratory, a plate discharge source has been used for \ce{CH3} radical production for several years now and, working with \ce{CH4} as a precursor, has proven very reliable.

\begin{acknowledgments}
Funding by the Deutsche Forschungsgemeinschaft (International Research Training Group 2079) is gratefully acknowledged. K.D. acknowledges financial support by the Fonds der Chemischen Industrie (FCI) for financial support through a Liebig fellowship. The study was also supported by a National Science and Engineering Research Discovery Grant in Canada and funds from the Canada Foundation for Innovation (CFI) for the Centre for Research on Ultra-Cold Systems (CRUCS) at UBC. The authors are thankful to Edvardas Narevicius (Weizmann Institute, Israel) for the loan of the DBD head.
\end{acknowledgments}


\providecommand{\latin}[1]{#1}
\makeatletter
\providecommand{\doi}
{\begingroup\let\do\@makeother\dospecials
	\catcode`\{=1 \catcode`\}=2\doi@aux}
\providecommand{\doi@aux}[1]{\endgroup\texttt{#1}}
\makeatother
\providecommand*\mcitethebibliography{\thebibliography}
\csname @ifundefined\endcsname{endmcitethebibliography}
{\let\endmcitethebibliography\endthebibliography}{}


\end{document}